\documentclass[english,aps,pra,twocolumn,showpacs]{revtex4}
\usepackage[T1]{fontenc}
\usepackage[latin9]{inputenc}
\usepackage{amsmath}
\usepackage{amssymb}
\usepackage{graphicx}

\usepackage{babel}
\begin{document}

\title{Generalized Holstein model for spin-dependent electron transfer reaction}

\author{Li-Ping Yang}

\affiliation{Institute of Theoretical Physics, Chinese Academy of Sciences, Beijing,
100190, China}

\author{Qing Ai}

\affiliation{Institute of Theoretical Physics, Chinese Academy of Sciences, Beijing,
100190, China}

\author{C. P. Sun}

\email{suncp@itp.ac.cn}

\homepage{http://power.itp.ac.cn/~suncp}

\affiliation{Institute of Theoretical Physics, Chinese Academy of Sciences, Beijing,
100190, China}

\date{\today}
\begin{abstract}
Some chemical reactions are described by electron transfer (ET) processes.
The underlying mechanism could be modeled as a polaron motion in the
molecular crystal\textemdash{}the Holstein model. By taking spin degrees
of freedom into consideration, we generalize the Holstein model (molecular
crystal model) to microscopically describe an ET chemical reaction.
In our model, the electron spins in the radical pair simultaneously
interact with a magnetic field and their nuclear-spin environments.
By virtue of the perturbation approach, we obtain the chemical reaction
rates for different initial states. It is discovered that the chemical
reaction rate of the triplet state demonstrates its dependence on
the direction of the magnetic field while the counterpart of the singlet
state does not. This difference is attributed to the explicit dependence
of the triplet state on the direction when the axis is rotated. Our
model may provide a possible candidate for the microscopic origin
of avian compass.
\end{abstract}


\maketitle

\section{Introduction}

Nowadays, it has been prevailing in both experimental and theoretical
explorations that quantum coherence effect due to the role of phase
in quantum superposition may exist in living processes. This essentially
implies that there may exist quantum coherence effect in chemical
reactions in some living processes, such as charge and energy transfer
in photosynthesis~\cite{Engel07,Lee07,Collini10,Yang_Sun10,Dong_Sun11}
and singlet-and-triplet transition in avian compass~\cite{Maeda08,Kominis09,derry,Schulten,Schulten-1,Schulten-2,Zapka09,Cai10,Cai11,Gauger11,Cai_Sun11}.

It has long been questioned how migratory birds can navigate to their
destination over hundreds of miles. One of the possible answers is
given by the radical pair mechanism~\cite{Schulten,Schulten-1,Schulten-2}.
Two unpaired electron spins in the radical pair are initially prepared
in the singlet state. Due to their interactions with the geomagnetic
field and their environmental nuclear spins, the election spins coherently
transit between the singlet and triplet states. Since the singlet
and  triplet states could result in different products of chemical
reactions, the direction and magnitude of the geomagnetic field determine
the relative yields of two distinct products. By sensing the information
incorporated in the final products of the chemical reactions in their
retinas, the birds can find their way to their destination. Therefore,
the quantum coherence underlies in the avian compass since the singlet
and triplet spin states correspond to different types of quantum entanglement.
Ever since it was proposed a decade ago, the radical-pair-mechanism-based
avian compass has been in favor by a series of biological and chemical
experiments~\cite{Maeda08,Kominis09}.

In this hypothesis, the nuclear spins play a crucial role because
there would be no coherent transition between the singlet and the
triplet states if there were no nuclear spins~\cite{Schulten-2}.
Previous studies mainly concentrated on the nuclear-spin environment
without inter-coupling~\cite{Cai10,Cai11,Gauger11}. Mostly recently,
by taking into account the inter-coupling of the nuclear spins, we
studied a special avian compass model with the nuclear environments
modeled by an Ising model in a transverse field~\cite{Cai_Sun11}.
The rationality of this model lies in the fact that the weak inter-nuclear-spin
coupling is comparable with the Zeeman energy splitting induced by
the weal geomagnetic field. It was discovered that the quantum criticality
in the environments enhances the sensitivity of magneto-reception.
On the other hand, although various master-equation approaches were
proposed to deal with such spin-dependent chemical reactions in the
avian compass~\cite{Kominis09}, the underlying physical mechanism
is still missing in studying the quantum coherence with microscopic
models. Thus, it is urgent to propose appropriate microscopic models
for different kinds of chemical reactions to make the quantum coherence
effect in those processes better understood. A case in point is the
Holstein's molecular crystal model, which is also regarded as a microscopic
model of chemical reactions with electron transfer (ET)~\cite{Schatz}.

The Holstein model was originally proposed to characterize the vibration-assisted
ET in one-electron molecular crystal~\cite{Schatz}. Here, in order
to describe the chemical reaction of spin dependence as well as direction
dependence, the Holstein model is generalized to incorporate the degrees
of freedom of spin to make electrons naturally interact with a magnetic
field. Additionally, due to the presence of the nuclear-spin environments
surrounding the electron spins, there would be coherent transition
between the singlet and triplet states of the two electron spins.
In contrast to the previous investigation using anisotropic hyperfine
coupling~\cite{Schulten-2}, the hyperfine interaction between the
electron spin and its nuclear environment is isotropic in our model.
Based on this generalized model, we calculate the chemical reaction
rates of the singlet and triplet states of the electron spins. Here,
the chemical reaction rate is determined by the transition rate of
one electron in a localized molecular orbit to another at a distance.
It is discovered that the reaction rate of the triplet state sensitively
responses to the variation of the direction of the magnetic field
with respect to the polarization of two electron spins. On the contrary,
the chemical reaction of the singlet state does not demonstrate such
dependence on the direction of the magnetic field. The above results
are attributed to the invariance of the singlet state under the rotation
of the system around $y$-axis, while the triplet one will be changed
along with the rotation according to irreducible tensor of $SO(3)$
group. Therefore, our proposed model may serve as a microscopic origin
for the chemical reaction in the avian compass.

In the next section, we generalize the Holstein model to incorporate
the electron spin degrees. In Sec.~III, we consider a general case
with an external magnetic field and nuclear-spin environments. In
Sec.~IV, we study the dynamic evolution of the radical pair and obtain
the chemical reaction rates for different initial states. Finally,
we summarize our main results in the Conclusion. Furthermore, we show
the detailed calculations for the chemical reaction probability, the
chemical reaction rate and the transition probability from the triplet
state to the singlet state in Appendix A and B respectively.

\section{Generalized Holstein model}

\begin{figure}
\includegraphics[width=8cm]{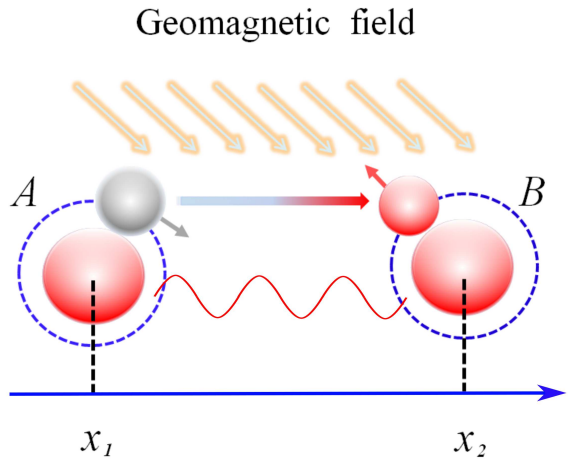}

\caption{(color online). Schematic diagram for the generalized Holstein model.
It is composed of a radical pair with nuclear spins. Two electrons
are initially prepared in a correlated state, i.e., the singlet state
or triplet state, which can be interconverted by the hyperfine interaction
in combination with an external magnetic field. The chemical reaction
occurs once both electrons are in the same site due to the tunneling
effect. }
\end{figure}

Many chemical reactions are accompanied by ET, where the electron
is transfered from one site to another (Fig.~1). A very important
but simple quantum-mechanical model for ET reactions is the molecular
crystal model, which was originally developed by Holstein to describe
so-called polaron motion in narrow-band conductors~\cite{Holstein}
and then understood as a microscopic model in describing simple chemical
reactions \cite{Schatz}.

The model Hamiltonian~$H=H_{v}+H_{e}+H_{ev}$ is decomposed into
two parts, i.e., the ionic vibration part
\begin{eqnarray}
H_{v} & \!\!=\!\! & \frac{1}{2}\sum_{j=1}^{2}(-\frac{\hbar^{2}}{m_{j}}\frac{\partial^{2}}{\partial x_{j}^{2}}+m_{j}\omega_{j}^{2}x_{j}^{2})+\sqrt{m_{1}m_{2}}fx_{1}x_{2},
\end{eqnarray}
and the electron-phonon hybrid part
\begin{align}
H_{e}\!+\! H_{ev} & \!\!=\!\!\sum_{j,\alpha}\!\bar{\varepsilon}_{j}(x_{1},x_{2})c_{j\alpha}^{\dagger}c_{j\alpha}\!-\! J\sum_{\alpha}\!(c_{1\alpha}^{\dagger}c_{2\alpha}\!+\!\mathrm{h.c.}),
\end{align}
where $x_{j}$ is the displacement of the $j$th ion, $\omega_{j}$
the harmonic vibration frequency with the reduced mass $m_{j}$, $f$
the coupling constant of the two molecules, and $c_{j\alpha}(c_{j\alpha}^{\dagger})$
the fermionic annihilation (creation) operator of the electron at
molecule $j$ with spin $\alpha$. Since the orbital energy $\bar{\varepsilon}(x_{1},x_{2})$
is linearized as
\begin{equation}
\bar{\varepsilon}_{j}(x_{1},x_{2})\approx\varepsilon_{j}+\sum_{i}\left(\frac{\partial\overline{\varepsilon}_{j}}{\partial x_{i}}\right)x_{i},
\end{equation}
we explicitly obtain the electronic Hamiltonian
\begin{equation}
H_{e}=\sum_{j,\alpha}\varepsilon_{j}c_{j\alpha}^{\dagger}c_{j\alpha}-J\sum_{\alpha}(c_{1\alpha}^{\dagger}c_{2\alpha}+\mathrm{h.c.}),
\end{equation}
and the electron-vibration coupling
\begin{equation}
H_{ev}=-\frac{1}{2}\sum_{j,\alpha}A_{j}c_{j\alpha}^{\dagger}c_{j\alpha}x_{j}.
\end{equation}
Here, the molecular orbital energy $\varepsilon_{j}$ is spin-independent.
In the next section, we will consider a more general case with an
external magnetic field. The tunneling integral $J$ is assumed to
be independent of the displacement $x_{j}$ and~$A_{j}=\sqrt{2}\partial\overline{\varepsilon}_{j}/\partial x_{j}$
denotes the electron-vibration coupling, where the term~$\partial\overline{\varepsilon}_{j}/\partial x_{i}$
for $i\neq j$ is neglected because the molecular orbital energy of
the $j$th molecule changes negligibly when the displacement of the
$i$th molecule varies.

For simplicity, we assume two identical molecules, i.e., $m_{1}=m_{2}=m$,
$\omega_{1}=\omega_{2}$, and $A_{1}=A_{2}=A$. Choosing coordinates
$X=(x_{1}+x_{2})/\sqrt{2}$ and $x=(x_{1}-x_{2})/\sqrt{2}$, we decompose
the Hamiltonian $H=H_{r}+H_{c}$ into two decoupled parts, i.e., the
one for the motion of the center of mass
\begin{equation}
H_{c}=-\frac{\hbar^{2}}{2m}\frac{\partial^{2}}{\partial X^{2}}+\frac{1}{2}m\Omega^{2}X^{2}+AX
\end{equation}
 with $\Omega=\sqrt{\omega_{1}^{2}+f}$, and the other for the relative
motion\begin{widetext}

\begin{eqnarray}
H_{r} & = & \hbar\omega(b^{\dagger}b+\frac{1}{2})+\sum_{j,\alpha}\varepsilon_{j}c_{j\alpha}^{\dagger}c_{j\alpha}-J\sum_{\alpha}(c_{1\alpha}^{\dagger}c_{2\alpha}+\mathrm{h.c.})-\frac{1}{2}A\sqrt{\frac{\hbar}{2m\omega}}(b^{\dagger}+b)\sum_{\alpha}(c_{2\alpha}^{\dagger}c_{2\alpha}-c_{1\alpha}^{\dagger}c_{1\alpha}),
\end{eqnarray}
\end{widetext}where we have introduced the bosonic operators
\begin{equation}
b=\sqrt{\frac{m\omega}{2\hbar}}(x+\frac{\hbar}{m\omega}\frac{\partial}{\partial x}),\ b^{\dagger}=\sqrt{\frac{m\omega}{2\hbar}}(x-\frac{\hbar}{m\omega}\frac{\partial}{\partial x}),
\end{equation}
$\omega=\sqrt{\omega_{1}^{2}-f}$ is the effective frequency.

Next, we make the Van Vleck transformation, also called the polaron
transformation~\cite{van Vleck,van Vleck-1},
\begin{equation}
\check{H}=e^{is}He^{-is}\label{van Vleck transformation}
\end{equation}
 for the above generalized Holstein model, where
\begin{equation}
s=-i\sum_{j,\alpha}\phi_{j}(b^{\dagger}-b)c_{j\alpha}^{\dagger}c_{j\alpha}
\end{equation}
is the transformation kernel and
\begin{equation}
\phi_{1}=-\phi_{2}=\frac{A}{2\hbar\omega}\sqrt{\frac{\hbar}{2m\omega}}\equiv\phi.
\end{equation}
 Thus we can formally decouple the degrees of freedom of electron
and vibration, obtaining $\check{H}_{r}=\check{H}_{r}^{(0)}+\check{H}_{r}^{(1)}$,
with
\begin{eqnarray}
\check{H}_{r}^{(0)} & \!=\! & \hbar\omega b^{\dagger}b\!+\!\sum_{j\alpha}\!\varepsilon_{j}c_{j\alpha}^{\dagger}c_{j\alpha}\!-\!\hbar\omega\!\left(\!\sum_{j\alpha}\!\phi_{j}c_{j\alpha}^{\dagger}c_{j\alpha}\!\right)^{2}\!\!\!\!,\label{eq:Holstein}
\end{eqnarray}
and
\begin{equation}
\check{H}_{r}^{(1)}\!=\!-J\sum_{\alpha}\!(c_{1\alpha}^{\dagger}c_{2\alpha}e^{2\phi(b^{\dagger}-b)}\!+\! c_{2\alpha}^{\dagger}c_{1\alpha}e^{-2\phi(b^{\dagger}-b)}).\label{Holstein1}
\end{equation}

This molecular crystal Hamiltonian~(\ref{eq:Holstein}) describes
the ET process for a two-local-orbit system. Here, we generalize the
Holstein model by taking into consideration the degrees of freedom
of the electron spins. Up to now, the above generalization seems to
be trivial, since we could totally separate the spin and orbital degrees
of freedom. However, when a local external magnetic field is applied
to the radical pair to form asymmetric couplings to the two electron
spins, the spin-orbit coupling is induced. In this case, a spin-dependent
ET process takes place. These asymmetric couplings can also be implemented
by coupling to their nuclear-spin environments.

\section{Spin molecular crystal in magnetic field and nuclear environment}

In the previous section, we described the generalized Holstein model,
with spin degree of freedom. In this section, on account of an external
magnetic field and nuclear-spin environments, we investigate how a
chemical reaction responses to its magnetic environment.

Choosing the polarization direction of the spin state as the $z$-direction,
we define the singlet state $\left|s\right\rangle $ and triplet state
$\left|t\right\rangle $ of the electron spins as
\begin{equation}
|s\rangle\!\!=\!\!\frac{1}{\sqrt{2}}\!\!\left(\left|\uparrow_{1}^{e}\downarrow_{2}^{e}\right\rangle \!-\!\left|\downarrow_{1}^{e}\uparrow_{2}^{e}\right\rangle \right)\!\!=\!\!\frac{1}{\sqrt{2}}\!\!\left(c_{1\uparrow}^{\dagger}c_{2\downarrow}^{\dagger}\!-\! c_{1\downarrow}^{\dagger}c_{2\uparrow}^{\dagger}\right)\!\left|0\right\rangle \!,
\end{equation}
and
\begin{equation}
|t\rangle\!=\!\frac{1}{\sqrt{2}}\!\left(\left|\uparrow_{1}^{e}\downarrow_{2}^{e}\right\rangle \!+\!\left|\downarrow_{1}^{e}\uparrow_{2}^{e}\right\rangle \right)\!=\!\frac{1}{\sqrt{2}}\!\left(c_{1\uparrow}^{\dagger}c_{2\downarrow}^{\dagger}\!+\! c_{1\downarrow}^{\dagger}c_{2\uparrow}^{\dagger}\right)\!\left|0\right\rangle \!,
\end{equation}
respectively with $\left|0\right\rangle $ being the vacuum state.

\begin{figure}
\includegraphics[width=8cm]{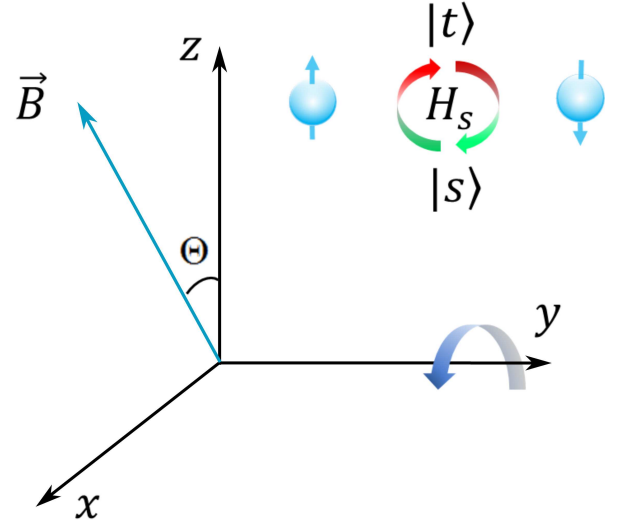}\caption{(color online). The inclination angle between the external geomagnetic
field and the $z$-direction is $\Theta$, and the coordinate system
is rotated around the $y$-axis with $\Theta$ to coincide with the
direction of the the magnetic field. The hyperfine coupling induced interconversion between
the singlet and triplet states is modulated by the direction of the magnetic field, i.e.,~$\Theta$.}
\end{figure}

In a simple case where the hyperfine couplings are isotropic, the
Hamiltonian, which describes the interaction between the electron
spins and their asymmetric magnetic environments (the magnetic field
plus the nuclear spins), reads
\begin{equation}
H_{s}=-\sum_{j=1}^{2}(\mu_{B}\vec{B}_{0}\cdot\hat{S}_{j}+g_{j}\hat{I}_{j}\cdot\hat{S}_{j}),
\end{equation}
where $\vec{B}_{0}=B_{0}(\sin\Theta,0,\cos\Theta)$ is the external
geomagnetic field with the inclination angle $\Theta$, $\mu_{B}$
the Bohr magneton, $\hat{S}_{j}=(S_{j}^{x},S_{j}^{y},S_{j}^{z})$
the Pauli operators for $j$th electron spin, $\hat{I}_{j}=(I_{j}^{x},I_{j}^{y},I_{j}^{z})$
the Pauli operators for $j$th nuclear spin, and $g_{j}$ the hyperfine
coupling constant between the $j$th electron spin and its environmental
nuclear spin.

Combining the relative vibration and spin Hamiltonians, the total
Hamiltonian for a spin-dependent ET reaction is obtained $H_{tot}=H_{r}+H_{s}$.
After a rotation around $y$-axis with the angle $\Theta$ (Fig.~2),
combined with the Van Vleck transformation defined in Eq.~(\ref{van Vleck transformation}),
we obtain $\tilde{H}_{tot}=\tilde{H}_{r}+\tilde{H}_{s}$, where the
relative vibration Hamiltonian $\tilde{H}_{r}=\check{H}_{r}$ is the
same as that given in Eqs.~(\ref{eq:Holstein}-\ref{Holstein1}),
but the Hamiltonian of the spin part is changed into

\begin{equation}
\tilde{H}_{s}=-\sum_{j}(\mu_{B}B_{0}S_{jz}+g_{j}\hat{I}_{j}\cdot\hat{S}_{j}).
\end{equation}
Meanwhile, we make the same rotation and transformation for the quantum
states of the whole system as in Eq.~(\ref{van Vleck transformation}).
Straightforwardly, after the combined transformation, the singlet
and triplet states read as $|\tilde{s}\rangle=|s\rangle$ and
\begin{eqnarray}
|\tilde{t}\rangle & = & \cos\Theta|t\rangle+\frac{1}{\sqrt{2}}\sin\Theta\!\left(\left|\uparrow_{1}^{e}\uparrow_{2}^{e}\right\rangle -\left|\downarrow_{1}^{e}\downarrow_{2}^{e}\right\rangle \right),
\end{eqnarray}
respectively.

It is obvious that the singlet state is not $\Theta$-dependent, while
the triplet state is. In the next section, we will study the dependence
of chemical reaction rate on the direction of the geomagnetic field.
As shown as follows, it is this explicit dependence on the direction
in the rotated triplet state that results in the variation of its
chemical reaction rate along with the changes of the magnetic field
direction.

\section{Magnetic direction controlling chemical reaction}

\begin{figure}
\includegraphics[width=8cm]{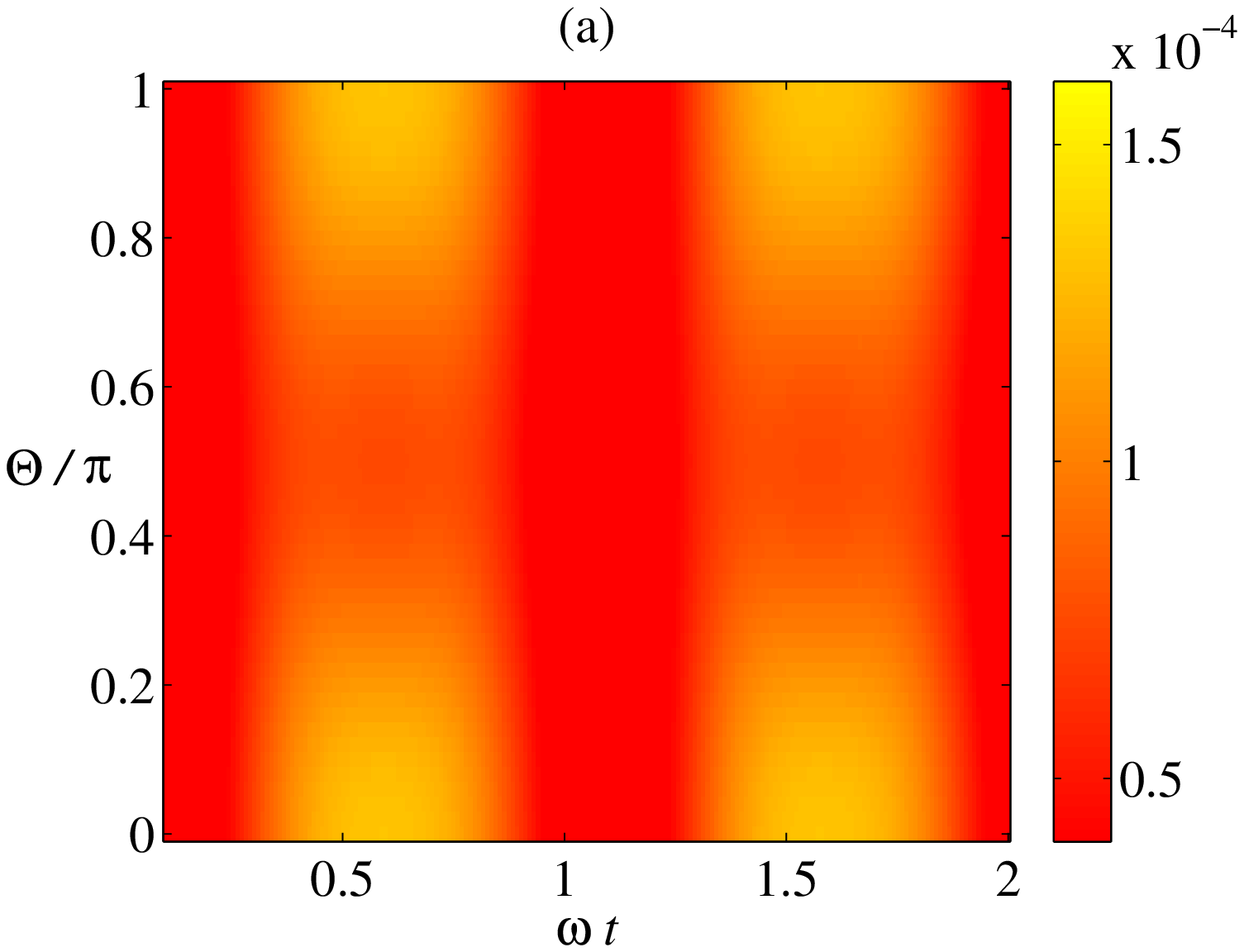}

\includegraphics[width=8cm]{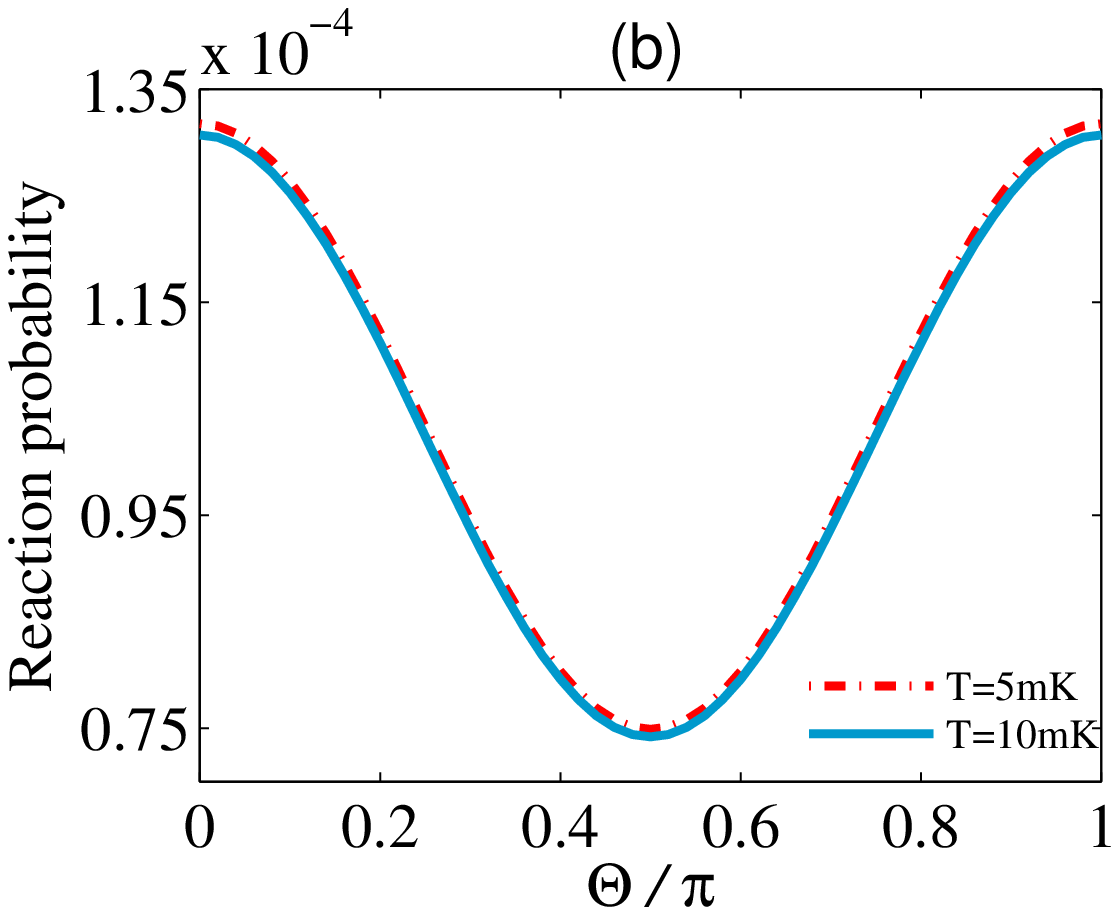}

\caption{(color online). $(a)$ Chemical reaction probability vs time and $\Theta$.
$(b)$ Chemical reaction probability for triplet state at a given
time $t=0.5/\omega$ for effective temperature $T=5mK$ (red dashed
line) and $T=10mK$ (blue solid line).}
\end{figure}

In this section, by means of the perturbation method, we analytically
obtain the probability for one electron to transfer to the other local
orbit to complete a chemical reaction. Assuming that at the initial
time the vibration and nuclear spins are both in thermal equilibrium
states and the electron spins are in the triplet state, the density
matrix of the whole system $\rho(0)=\rho_{v}\otimes\rho_{n}\otimes\rho_{t}$
includes three parts. The first part
\begin{equation}
\rho_{v}=\frac{1}{Z}\sum_{m=0}^{\infty}\exp\left(-\frac{m\hbar\omega}{k_{B}T}\right)\left|m\right\rangle \left\langle m\right|
\end{equation}
denotes the relative vibration of the molecules, where $Z=1/[1-\exp(-\hbar\omega/k_{B}T)]$,
$k_{B}$ the Boltzmann constant, and $T$ the temperature of the environment.
The second part is the density matrix of the nuclear spins. Since
$\mu_{n}B_{0}\ll k_{B}T$ with $\mu_{n}$ the nuclear magneton, the
nuclear spins are in the state
\begin{equation}
\rho_{n}=\frac{1}{4}\sum_{j=1}^{4}\left|\chi_{j}^{n}\right\rangle \left\langle \chi_{j}^{n}\right|,
\end{equation}
where $|\chi_{1}^{n}\rangle=\left|\downarrow_{1}^{n}\downarrow_{2}^{n}\right\rangle $,
$|\chi_{2}^{n}\rangle=\left|\downarrow_{1}^{n}\uparrow_{2}^{n}\right\rangle $,
$|\chi_{3}^{n}\rangle=\left|\uparrow_{1}^{n}\downarrow_{2}^{n}\right\rangle $,
and $|\chi_{4}^{n}\rangle=\left|\uparrow_{1}^{n}\uparrow_{2}^{n}\right\rangle $.
The last one
\begin{equation}
\rho_{t}=\left|t\right\rangle \left\langle t\right|
\end{equation}
describes the electrons.

Starting from the above initial state, we calculate the total ET reaction
probability of the triplet state (for the details please refer to
Appendix A)
\begin{equation}
P_{t}(\tau)=\frac{1}{4Z}\sum_{m,n=0}^{\infty}\sum_{j=1}^{4}\sum_{p=1}^{24}e^{-\beta m\hbar\omega}P_{jmnp}(\tau),\label{eq:transition probability}
\end{equation}
 where
\begin{equation}
P_{jmnp}(\tau)=\left|\sum_{q}c_{jmq}\tilde{H}_{np,mq}^{(1)}\frac{1-e^{i\omega_{np,mq}\tau}}{\hbar\omega_{np,mq}}\right|^{2},
\end{equation}
and~$c_{jmq}=\langle\psi_{mq}^{(0)}|\tilde{t}\rangle|\chi_{j}^{n}\rangle|m\rangle$
is the expanding coefficient. And the chemical reaction rate is determined
by the reaction probability per unite time in the long-time limit~\cite{Sakurai}
\begin{equation}
k_{t}=\frac{\partial}{\partial\tau}\lim_{\tau\rightarrow\infty}P_{t}(\tau),\label{eq:rate constant}
\end{equation}
where the explicit expression which displays direction dependence
is given in Appendix A.

To show the above results in an intuitive way, we turn to numerical
examples. We take the orbital energy difference~$\Delta=\varepsilon_{1}-\varepsilon_{2}=0.01\mathrm{eV}$
and the relative vibration frequency $\omega=10^{7}\mathrm{Hz}$.
We assume the tunneling integral $J=0.01\Delta$ and $\phi=0.2$.
The magnitude of the geomagnetic field is $B_{0}=50\mu\mathrm{T}$
and the hyperfine coupling constant is~$g_{1}=g_{2}=10^{-8}\mathrm{eV}$.
In the following calculations, we need to take a cut of the phonon
occupation number, defined by an effective temperature of the environment.
We find that the chemical reaction probability is not sensitive to
the temperature of the environment. For these parameters we numerically
calculate the transition probability for initial triplet state from
Eq.$~$(\ref{eq:transition probability}) as shown in Fig.~3.

Obviously, the ET probability displays its dependence on the angle
$\Theta.$ At a given time, the ET probability falls to its minimum
value when the magnetic field is perpendicular to $z$ direction,
while it reaches its maximum when the direction of the external field
is parallel to $z$ axis. Besides, the probability is symmetrical
about $\Theta=\pi/2$. When we come to the initial singlet state,
there is no such dependence on the angle $\Theta$. This is a reasonable
result for isotropic hyperfine coupling case since both the singlet
state and the transformed Hamiltonian do not explicitly depend on
the angle.

\begin{figure}
\includegraphics[width=8cm]{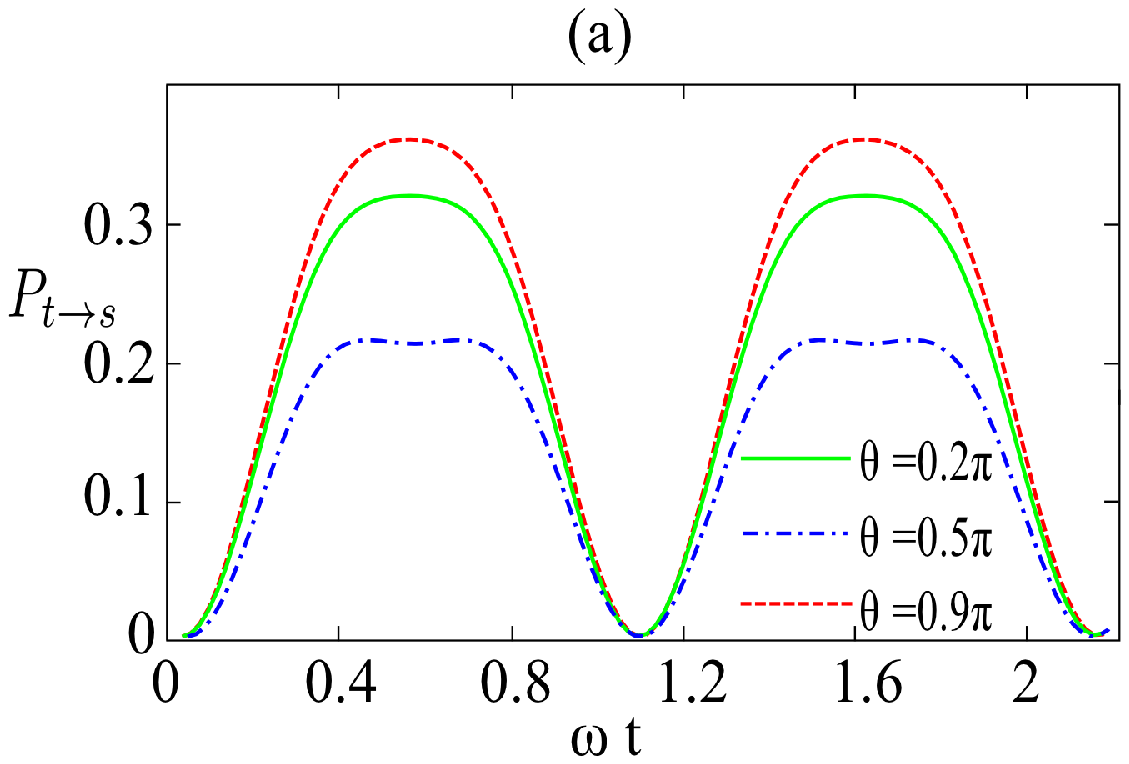}

\includegraphics[width=8cm]{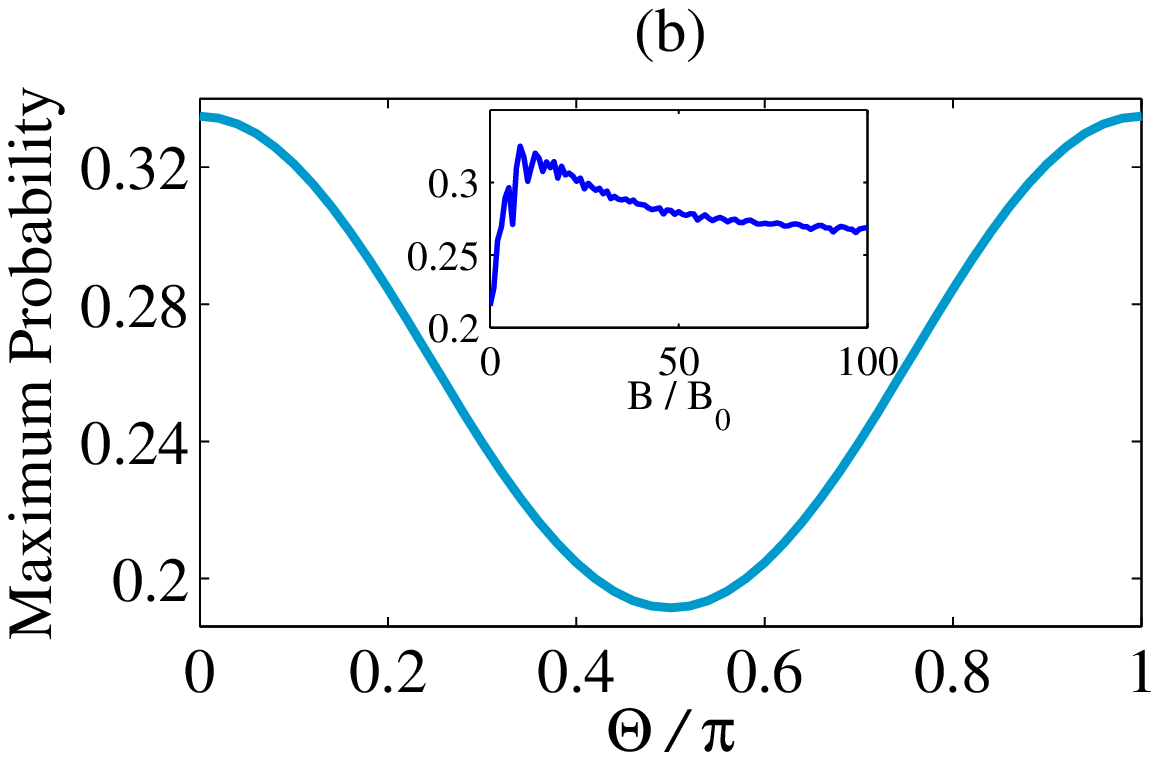}

\caption{(color online). Transition probability from the triplet state to singlet
state. $(a)$ Probability vs time for different angles. $(b)$ The
maximum probability varies with angle. The inset displays that the
maximum transition probability also changes with the magnitude of
the magnetic field.}
\end{figure}

On the other hand, the chemical reaction rate for triplet state would
vanish once there were no interaction between the electron spins and
their nuclear environments, i.e., $g_{j}=0$, while the ET reaction
happens when the electron spins are in the singlet state. This can
be seen from the fact that $H_{r}^{(1)}|t\rangle=0$ but $H_{r}^{(1)}|s\rangle\neq0$.
The coupling of electron spins to the nuclear-spin environments can
induce the transition from the triplet to the singlet states to make
the ET happen. Then the spatial ET leads to the chemical reaction.
In order to illustrate the mechanism of magnetic-direction-controlling
chemical reaction more clearly, we study the dynamic evolution of
the radical pair. Using the same parameters as those in Fig.~3, we
plot the transition probability from the triplet state to the singlet
state
\begin{equation}
P_{t\rightarrow s}=\mathrm{Tr}_{v,n,e}\left[\tilde{\rho}_{s}e^{-i\tilde{H}t/\hbar}\tilde{\rho}(0)e^{i\tilde{H}t/\hbar}\right]
\end{equation}
in Fig.~4. Due to the presence of the nuclear-spin environments,
there is periodical interconversion between the singlet and triplet
states. Besides, the amplitude of this periodical oscillation is adjusted
by the direction of the magnetic field. For one thing, the singlet-triplet
interconversion is induced by the hyperfine couplings, not the uniform
magnetic field, but it is modulated by the geomagnetic field. Our
conjecture, that the triplet state is converted to the singlet state
to complete the ET reaction, is confirmed by the same line shape of
the maximum conversion probability as that of the chemical reaction
probability of the triplet state. Besides, as shown in the inset of
Fig.~4(b), with the increasing of the magnitude of the magnetic field,
the maximum of the transition probability increase at first for a
given angle $\Theta=0.1\pi$ and finally decreases into a stable value.

\section{Conclusion}

On account of electron spin degrees of freedom, we study the effect
of the direction of the magnetic field on the chemical reaction by
generalizing the Holstein model. By means of the perturbation approach,
we obtain the ET reaction probability and chemical reaction rate of
the singlet and triplet states. The chemical reaction rate of the
triplet state displays its sensitive dependence on the direction of
the magnetic field in contrast to the counterpart of the singlet state.
We demonstrate that the triplet state indirectly participates in the
chemical reaction. It must be converted to the singlet state by the
hyperfine coupling between electrons and nuclear spins to take part
in the ET reaction. We emphasize that the hyperfine couplings are
isotropic in our model which are different from the anisotropic ones
in the previous study~\cite{Schulten-2}. With the above comprehensive
consideration, it could be concluded that our model may serve as a
possible microscopic origin for the avian compass.

\section*{Acknowledgement }

We thank Hui Dong, Cheng-Yun Cai, and Da Zhi Xu for helpful discussion.
This work is supported by National Natural Science Foundation of China
under Grant Nos. 10935010 and 11074261.

\appendix

\section{Chemical reaction rate}

In this appendix, we calculate the ET reaction probability to the
first order.

First of all, the Hamiltonian of the electron spin-$j$ part $\tilde{H}_{s}^{(j)}$
is diagonalized with eigenstates
\begin{equation}
\left|e_{1}^{(j)}\right\rangle =\left|\downarrow_{j}^{e}\right\rangle \left|\downarrow_{j}^{n}\right\rangle ,
\end{equation}

\begin{equation}
\left|e_{2}^{(j)}\right\rangle =\cos\frac{\theta_{j}}{2}\left|\downarrow_{j}^{e}\right\rangle \left|\uparrow_{j}^{n}\right\rangle -\sin\frac{\theta_{j}}{2}\left|\uparrow_{j}^{e}\right\rangle \left|\downarrow_{j}^{n}\right\rangle ,
\end{equation}

\begin{equation}
\left|e_{3}^{(j)}\right\rangle =\sin\frac{\theta_{j}}{2}\left|\downarrow_{j}^{e}\right\rangle \left|\uparrow_{j}^{n}\right\rangle +\cos\frac{\theta_{j}}{2}\left|\uparrow_{j}^{e}\right\rangle \left|\downarrow_{j}^{n}\right\rangle ,
\end{equation}

\begin{equation}
\left|e_{4}^{(j)}\right\rangle =\left|\uparrow_{j}^{e}\right\rangle \left|\uparrow_{j}^{n}\right\rangle ,
\end{equation}
and the corresponding eigenvalues
\begin{eqnarray}
e_{1}^{(j)} & = & \mu_{B}B_{0}-g_{j},\\
e_{2}^{(j)} & = & g_{j}+\sqrt{\mu_{B}^{2}B_{0}^{2}+4g_{j}^{2}},\\
e_{3}^{(j)} & = & g_{j}-\sqrt{\mu_{B}^{2}B_{0}^{2}+4g_{j}^{2}},\\
e_{4}^{(j)} & = & -\mu_{B}B_{0}-g_{j}.
\end{eqnarray}
Here, the mixing angles is defined as
\begin{equation}
\theta_{j}=\tan^{-1}\left(\frac{2g_{j}}{\mu_{B}B_{0}}\right).
\end{equation}
Straightforwardly, we explicitly calculate the eigenstates and eigenvalues
of the total Hamiltonian for the spins $\tilde{H}_{s}=\sum_{j}\tilde{H}_{s}^{(j)}$,
listed in Table~\mbox{I}.

\begin{table}
\begin{tabular}{|c|c|}
\hline
Eigenstate & Eigenvalue\tabularnewline
\hline
$\left|\varphi_{1}\right\rangle =|e_{1}^{(1)}\rangle\otimes|e_{1}^{(2)}\rangle$ & $E_{s1}=e_{1}^{(1)}+e_{1}^{(2)}$\tabularnewline
\hline
$\left|\varphi_{2}\right\rangle =|e_{1}^{(1)}\rangle\otimes|e_{2}^{(2)}\rangle$ & $E_{s2}=e_{1}^{(1)}+e_{2}^{(2)}$\tabularnewline
\hline
$\left|\varphi_{3}\right\rangle =|e_{1}^{(1)}\rangle\otimes|e_{3}^{(2)}\rangle$ & $E_{s3}=e_{1}^{(1)}+e_{3}^{(2)}$\tabularnewline
\hline
$\left|\varphi_{4}\right\rangle =|e_{1}^{(1)}\rangle\otimes|e_{4}^{(2)}\rangle$ & $E_{s4}=e_{1}^{(1)}+e_{4}^{(2)}$\tabularnewline
\hline
$\left|\varphi_{5}\right\rangle =|e_{2}^{(1)}\rangle\otimes|e_{1}^{(2)}\rangle$ & $E_{s5}=e_{2}^{(1)}+e_{1}^{(2)}$\tabularnewline
\hline
$\left|\varphi_{6}\right\rangle =|e_{2}^{(1)}\rangle\otimes|e_{2}^{(2)}\rangle$ & $E_{s6}=e_{2}^{(1)}+e_{2}^{(2)}$\tabularnewline
\hline
$\left|\varphi_{7}\right\rangle =|e_{2}^{(1)}\rangle\otimes|e_{3}^{(2)}\rangle$ & $E_{s7}=e_{2}^{(1)}+e_{3}^{(2)}$\tabularnewline
\hline
$\left|\varphi_{8}\right\rangle =|e_{2}^{(1)}\rangle\otimes|e_{4}^{(2)}\rangle$ & $E_{s8}=e_{2}^{(1)}+e_{4}^{(2)}$\tabularnewline
\hline
$\left|\varphi_{9}\right\rangle =|e_{3}^{(1)}\rangle\otimes|e_{1}^{(2)}\rangle$ & $E_{s9}=e_{3}^{(1)}+e_{1}^{(2)}$\tabularnewline
\hline
$\left|\varphi_{10}\right\rangle =|e_{3}^{(1)}\rangle\otimes|e_{2}^{(2)}\rangle$ & $E_{s10}=e_{3}^{(1)}+e_{2}^{(2)}$\tabularnewline
\hline
$\left|\varphi_{11}\right\rangle =|e_{3}^{(1)}\rangle\otimes|e_{3}^{(2)}\rangle$ & $E_{s11}=e_{3}^{(1)}+e_{3}^{(2)}$\tabularnewline
\hline
$\left|\varphi_{12}\right\rangle =|e_{3}^{(1)}\rangle\otimes|e_{4}^{(2)}\rangle$ & $E_{s12}=e_{3}^{(1)}+e_{4}^{(2)}$\tabularnewline
\hline
$\left|\varphi_{13}\right\rangle =|e_{4}^{(1)}\rangle\otimes|e_{1}^{(2)}\rangle$ & $E_{s13}=e_{4}^{(1)}+e_{1}^{(2)}$\tabularnewline
\hline
$\left|\varphi_{14}\right\rangle =|e_{4}^{(1)}\rangle\otimes|e_{2}^{(2)}\rangle$ & $E_{s14}=e_{4}^{(1)}+e_{2}^{(2)}$\tabularnewline
\hline
$\left|\varphi_{15}\right\rangle =|e_{4}^{(1)}\rangle\otimes|e_{3}^{(2)}\rangle$ & $E_{s15}=e_{4}^{(1)}+e_{3}^{(2)}$\tabularnewline
\hline
$\left|\varphi_{16}\right\rangle =|e_{4}^{(1)}\rangle\otimes|e_{4}^{(2)}\rangle$ & $E_{s16}=e_{4}^{(1)}+e_{4}^{(2)}$\tabularnewline
\hline
$\left|\varphi_{17}\right\rangle =\left|\uparrow_{1}^{e}\downarrow_{1}^{e}\right\rangle \otimes\left|\downarrow_{1}^{n}\downarrow_{2}^{n}\right\rangle $ & $E_{s17}=0$\tabularnewline
\hline
$\left|\varphi_{18}\right\rangle =\left|\uparrow_{1}^{e}\downarrow_{1}^{e}\right\rangle \otimes\left|\downarrow_{1}^{n}\uparrow_{2}^{n}\right\rangle $ & $E_{s18}=0$\tabularnewline
\hline
$\left|\varphi_{19}\right\rangle =\left|\uparrow_{1}^{e}\downarrow_{1}^{e}\right\rangle \otimes\left|\uparrow_{1}^{n}\downarrow_{2}^{n}\right\rangle $ & $E_{s19}=0$\tabularnewline
\hline
$\left|\varphi_{20}\right\rangle =\left|\uparrow_{1}^{e}\downarrow_{1}^{e}\right\rangle \otimes\left|\uparrow_{1}^{n}\uparrow_{2}^{n}\right\rangle $ & $E_{s20}=0$\tabularnewline
\hline
$\left|\varphi_{21}\right\rangle =\left|\uparrow_{2}^{e}\downarrow_{2}^{e}\right\rangle \otimes\left|\downarrow_{1}^{n}\downarrow_{2}^{n}\right\rangle $ & $E_{s21}=0$\tabularnewline
\hline
$\left|\varphi_{22}\right\rangle =\left|\uparrow_{2}^{e}\downarrow_{2}^{e}\right\rangle \otimes\left|\downarrow_{1}^{n}\uparrow_{2}^{n}\right\rangle $ & $E_{s21}=0$\tabularnewline
\hline
$\left|\varphi_{23}\right\rangle =\left|\uparrow_{2}^{e}\downarrow_{2}^{e}\right\rangle \otimes\left|\uparrow_{1}^{n}\downarrow_{2}^{n}\right\rangle $ & $E_{s22}=0$\tabularnewline
\hline
$\left|\varphi_{24}\right\rangle =\left|\uparrow_{2}^{e}\downarrow_{2}^{e}\right\rangle \otimes\left|\uparrow_{1}^{n}\uparrow_{2}^{n}\right\rangle $ & $E_{s24}=0$\tabularnewline
\hline
\end{tabular}

\caption{All 24 eigenstates and eigenvalues of $\tilde{H}_{s}$, where the
two electrons are located in two distant orbits respectively for the
first 16 eigenstates, while for the other $8$ eigenstates, both electrons
are in the same site.}
\end{table}

\begin{table*}
\begin{tabular}{|c|c|}
\hline
Coefficient & Explicit expression\tabularnewline
\hline
$R_{1,mn,21}$ & %
\begin{tabular}{c}
$\cos^{2}\Theta[\sin^{2}\frac{\theta_{2}}{2}\sin^{2}\frac{\theta_{2}}{2}\delta\left(E_{n,21}^{(0)}-E_{m,2}^{(0)}\right)+\cos^{2}\frac{\theta_{2}}{2}\cos^{2}\frac{\theta_{2}}{2}\delta\left(E_{n,21}^{(0)}-E_{m,3}^{(0)}\right)$\tabularnewline
$+\sin^{2}\frac{\theta_{1}}{2}\sin^{2}\frac{\theta_{1}}{2}\delta\left(E_{n,21}^{(0)}-E_{m,5}^{(0)}\right)+\cos^{2}\frac{\theta_{1}}{2}\cos^{2}\frac{\theta_{1}}{2}\delta\left(E_{n,21}^{(0)}-E_{m,9}^{(0)}\right)]$\tabularnewline
\end{tabular}\tabularnewline
\hline
$R_{1,mn,22}$ & %
\begin{tabular}{c}
$\frac{1}{4}\sin^{2}\Theta[\sin^{4}\frac{\theta_{1}}{2}\sin^{2}\theta_{2}\delta\left(E_{n,22}^{(0)}-E_{m,6}^{(0)}\right)+\sin^{4}\frac{\theta_{1}}{2}\sin^{2}\theta_{2}\delta\left(E_{n,22}^{(0)}-E_{m,7}^{(0)}\right)$\tabularnewline
$+\cos^{4}\frac{\theta_{1}}{2}\sin^{2}\theta_{2}\delta\left(E_{n,22}^{(0)}-E_{m,10}^{(0)}\right)+\cos^{4}\frac{\theta_{1}}{2}\sin^{2}\theta_{2}\delta\left(E_{n,22}^{(0)}-E_{m,11}^{(0)}\right)]$\tabularnewline
\end{tabular}\tabularnewline
\hline
$R_{1,mn,23}$ & %
\begin{tabular}{c}
$\frac{1}{4}\sin^{2}\Theta[\sin^{2}\theta_{1}\sin^{4}\frac{\theta_{2}}{2}\delta\left(E_{n,23}^{(0)}-E_{m,6}^{(0)}\right)+\sin^{2}\theta_{1}\cos^{4}\frac{\theta_{2}}{2}\delta\left(E_{n,23}^{(0)}-E_{m,7}^{(0)}\right)$\tabularnewline
$+\sin^{2}\theta_{1}\sin^{4}\frac{\theta_{2}}{2}\delta\left(E_{n,23}^{(0)}-E_{m,10}^{(0)}\right)+\sin^{2}\theta_{1}\cos^{4}\frac{\theta_{2}}{2}\delta\left(E_{n,23}^{(0)}-E_{m,11}^{(0)}\right)]$\tabularnewline
\end{tabular}\tabularnewline
\hline
$R_{1,mn,24}$ & $0$\tabularnewline
\hline
$R_{2,mn,21}$ & $\sin^{2}\Theta[\cos^{2}\frac{\theta_{2}}{2}\sin^{2}\frac{\theta_{2}}{2}\delta\left(E_{n,21}^{(0)}-E_{m,2}^{(0)}\right)+\sin^{2}\frac{\theta_{2}}{2}\cos^{2}\frac{\theta_{2}}{2}\delta\left(E_{n,21}^{(0)}-E_{m,3}^{(0)}\right)]$\tabularnewline
\hline
$R_{2,mn,22}$ & %
\begin{tabular}{c}
$\cos^{2}\Theta[\delta\left(E_{n,22}^{(0)}-E_{m,4}^{(0)}\right)+\sin^{4}\frac{\theta_{1}}{2}\cos^{4}\frac{\theta_{2}}{2}\delta\left(E_{n,22}^{(0)}-E_{m,6}^{(0)}\right)+\sin^{4}\frac{\theta_{1}}{2}\sin^{4}\frac{\theta_{2}}{2}\delta\left(E_{n,22}^{(0)}-E_{m,7}^{(0)}\right)$\tabularnewline
$+\cos^{4}\frac{\theta_{1}}{2}\cos^{4}\frac{\theta_{2}}{2}\delta\left(E_{n,22}^{(0)}-E_{m,10}^{(0)}\right)+\cos^{4}\frac{\theta_{1}}{2}\sin^{4}\frac{\theta_{2}}{2}\delta\left(E_{n,22}^{(0)}-E_{m,11}^{(0)}\right)]$\tabularnewline
\end{tabular}\tabularnewline
\hline
$R_{2,nm,23}$ & %
\begin{tabular}{c}
$\frac{1}{16}\cos^{2}\Theta[\sin^{2}\theta_{1}\sin^{2}\theta_{2}\delta\left(E_{n,23}^{(0)}-E_{m,6}^{(0)}\right)+\sin^{2}\theta_{1}\sin^{2}\theta_{2}\delta\left(E_{n,23}^{(0)}-E_{m,7}^{(0)}\right)$\tabularnewline
$+\sin^{2}\theta_{1}\sin^{2}\theta_{2}\delta\left(E_{n,23}^{(0)}-E_{m,10}^{(0)}\right)+\sin^{2}\theta_{1}\sin^{2}\theta_{2}\delta\left(E_{n,23}^{(0)}-E_{m,11}^{(0)}\right)]$\tabularnewline
\end{tabular}\tabularnewline
\hline
$R_{2,nm,24}$ & $\sin^{2}\Theta[\sin^{2}\frac{\theta_{1}}{2}\cos^{2}\frac{\theta_{1}}{2}\delta\left(E_{n,24}^{(0)}-E_{m,8}^{(0)}\right)+\cos^{2}\frac{\theta_{1}}{2}\sin^{2}\frac{\theta_{1}}{2}\delta\left(E_{n,24}^{(0)}-E_{m,12}^{(0)}\right)]$\tabularnewline
\hline
$R_{3,mn,21}$ & $\sin^{2}\Theta[\sin^{2}\frac{\theta_{1}}{2}\cos^{2}\frac{\theta_{1}}{2}\delta\left(E_{n,21}^{(0)}-E_{m,5}^{(0)}\right)+\sin^{2}\frac{\theta_{1}}{2}\cos^{2}\frac{\theta_{1}}{2}\delta\left(E_{n,21}^{(0)}-E_{m,9}^{(0)}\right)]$\tabularnewline
\hline
$R_{3,mn,22}$ & %
\begin{tabular}{c}
$\frac{1}{16}\cos^{2}\Theta[\sin^{2}\theta_{1}\sin^{2}\theta_{2}\delta\left(E_{n,22}^{(0)}-E_{m,6}^{(0)}\right)+\sin^{2}\theta_{1}\sin^{2}\theta_{2}\delta\left(E_{n,22}^{(0)}-E_{m,7}^{(0)}\right)$\tabularnewline
$+\sin^{2}\theta_{1}\sin^{2}\theta_{2}\delta\left(E_{n,22}^{(0)}-E_{m,10}^{(0)}\right)+\sin^{2}\theta_{1}\sin^{2}\theta_{2}\delta\left(E_{n,22}^{(0)}-E_{m,11}^{(0)}\right)]$\tabularnewline
\end{tabular}\tabularnewline
\hline
$R_{3,mn,23}$ & %
\begin{tabular}{c}
$\cos^{2}\Theta[\delta\left(E_{n,23}^{(0)}-E_{m,11}^{(0)}\right)+\cos^{4}\frac{\theta_{1}}{2}\sin^{4}\frac{\theta_{2}}{2}\delta\left(E_{n,23}^{(0)}-E_{m,6}^{(0)}\right)+\cos^{4}\frac{\theta_{1}}{2}\cos^{4}\frac{\theta_{2}}{2}\delta\left(E_{n,23}^{(0)}-E_{m,7}^{(0)}\right)$\tabularnewline
$+\sin^{4}\frac{\theta_{1}}{2}\sin^{4}\frac{\theta_{2}}{2}\delta\left(E_{n,23}^{(0)}-E_{m,10}^{(0)}\right)+\sin^{4}\frac{\theta_{1}}{2}\cos^{4}\frac{\theta_{2}}{2}\delta\left(E_{n,23}^{(0)}-E_{m,11}^{(0)}\right)]$\tabularnewline
\end{tabular}\tabularnewline
\hline
$R_{3,mn,24}$ & %
\begin{tabular}{c}
$\sin^{2}\Theta[\sin^{2}\frac{\theta_{2}}{2}\cos^{2}\frac{\theta_{2}}{2}\delta\left(E_{n,24}^{(0)}-E_{m,14}^{(0)}\right)+\sin^{2}\frac{\theta_{2}}{2}\cos^{2}\frac{\theta_{2}}{2}\delta\left(E_{n,24}^{(0)}-E_{m,15}^{(0)}\right)]$\tabularnewline
\end{tabular}\tabularnewline
\hline
$R_{4,mn,21}$ & $0$\tabularnewline
\hline
$R_{4,mn,22}$ & %
\begin{tabular}{c}
$\frac{\sin^{2}\Theta}{4}[\sin^{2}\theta_{1}\cos^{4}\frac{\theta_{2}}{2}\delta\left(E_{n,22}^{(0)}-E_{m,6}^{(0)}\right)+\sin^{2}\theta_{1}\sin^{4}\frac{\theta_{2}}{2}\delta\left(E_{n,22}^{(0)}-E_{m,7}^{(0)}\right)$\tabularnewline
$+\sin^{2}\theta_{1}\cos^{4}\frac{\theta_{2}}{2}\delta\left(E_{n,22}^{(0)}-E_{m,10}^{(0)}\right)+\sin^{2}\theta_{1}\cos^{2}\frac{\theta_{2}}{2}\delta\left(E_{n,22}^{(0)}-E_{m,11}^{(0)}\right)]$\tabularnewline
\end{tabular}\tabularnewline
\hline
$R_{4,mn,23}$ & %
\begin{tabular}{c}
$\frac{\sin^{2}\Theta}{4}[\cos^{4}\frac{\theta_{1}}{2}\sin^{2}\theta_{2}\delta\left(E_{n,23}^{(0)}-E_{m,6}^{(0)}\right)+\cos^{4}\frac{\theta_{1}}{2}\sin^{2}\theta_{2}\delta\left(E_{n,23}^{(0)}-E_{m,7}^{(0)}\right)$\tabularnewline
$+\sin^{4}\frac{\theta_{1}}{2}\sin^{2}\theta_{2}\delta\left(E_{n,23}^{(0)}-E_{m,10}^{(0)}\right)+\sin^{4}\frac{\theta_{1}}{2}\sin^{2}\theta_{2}\delta\left(E_{n,23}^{(0)}-E_{m,11}^{(0)}\right)]$\tabularnewline
\end{tabular}\tabularnewline
\hline
$R_{4,mn,24}$ & %
\begin{tabular}{c}
$\cos^{2}\Theta[\cos^{4}\frac{\theta_{1}}{2}\delta\left(E_{n,24}^{(0)}-E_{m,8}^{(0)}\right)+\sin^{4}\frac{\theta_{1}}{2}\delta\left(E_{n,24}^{(0)}-E_{m,11}^{(0)}\right)$\tabularnewline
$+\cos^{4}\frac{\theta_{2}}{2}\delta\left(E_{n,24}^{(0)}-E_{m,14}^{(0)}\right)+\sin^{2}\frac{\theta_{2}}{2}\delta\left(E_{n,24}^{(0)}-E_{m,15}^{(0)}\right)]$\tabularnewline
\end{tabular}\tabularnewline
\hline
\end{tabular}

\caption{Coefficients for the chemical reaction rate of the triplet state.}
\end{table*}

\begin{table*}
\begin{tabular}{|c|c|}
\hline
Coefficient & Explicit expression\tabularnewline
\hline
$D_{m,1,1}$ & %
\begin{tabular}{c}
$\frac{1}{2}[-\sin\frac{\theta_{2}}{2}\sin\frac{\theta_{2}}{2}\cos\Theta e^{-iE_{m,2}t/\hbar}-\cos\frac{\theta_{2}}{2}\cos\frac{\theta_{2}}{2}\cos\Theta e^{-iE_{m,3}t/\hbar}$\tabularnewline
$+\sin\frac{\theta_{1}}{2}\sin\frac{\theta_{1}}{2}\cos\Theta e^{-iE_{m,5}t/\hbar}+\cos\frac{\theta_{1}}{2}\cos\frac{\theta_{1}}{2}\cos\Theta e^{-iE_{m,9}t/\hbar}]$\tabularnewline
\end{tabular}\tabularnewline
\hline
$D_{m,1,2}$ & %
\begin{tabular}{c}
$\frac{1}{4}[-\sin^{2}\frac{\theta_{1}}{2}\sin\theta_{2}\sin\Theta e^{-iE_{m,6}t/\hbar}+\sin^{2}\frac{\theta_{1}}{2}\sin\theta_{2}\sin\Theta e^{-iE_{m,7}t/\hbar}$\tabularnewline
$-\cos^{2}\frac{\theta_{1}}{2}\sin\theta_{2}\sin\Theta e^{-iE_{m,10}t/\hbar}+\cos^{2}\frac{\theta_{1}}{2}\sin\theta_{2}\sin\Theta e^{-iE_{m,11}t/\hbar}]$\tabularnewline
\end{tabular}\tabularnewline
\hline
$D_{m,1,3}$ & %
\begin{tabular}{c}
$\frac{1}{4}[\sin\theta_{1}\sin^{2}\frac{\theta_{2}}{2}\sin\Theta e^{-iE_{m,6}t/\hbar}+\sin\theta_{1}\cos^{2}\frac{\theta_{2}}{2}\sin\Theta e^{-iE_{m,7}t/\hbar}$\tabularnewline
$-\sin\theta_{1}\sin^{2}\frac{\theta_{2}}{2}\sin\Theta e^{-iE_{m,10}t/\hbar}-\sin\theta_{1}\cos^{2}\frac{\theta_{2}}{2}\sin\Theta e^{-iE_{m,11}t/\hbar}]$\tabularnewline
\end{tabular}\tabularnewline
\hline
$D_{m,1,4}$ & $0$\tabularnewline
\hline
$D_{m,2,1}$ & $\frac{1}{2}[-\sin\frac{\theta_{2}}{2}\cos\frac{\theta_{2}}{2}\sin\Theta e^{-iE_{m,2}t/\hbar}+\cos\frac{\theta_{2}}{2}\sin\frac{\theta_{2}}{2}\sin\Theta e^{-iE_{m,3}t/\hbar}]$\tabularnewline
\hline
$D_{m,2,2}$ & %
\begin{tabular}{c}
$\frac{1}{2}[-\cos\Theta e^{-iE_{m,4}t/\hbar}+\sin^{2}\frac{\theta_{1}}{2}\cos^{2}\frac{\theta_{2}}{2}\cos\Theta e^{-iE_{m,6}t/\hbar}+\sin^{2}\frac{\theta_{1}}{2}\sin^{2}\frac{\theta_{2}}{2}\cos\Theta e^{-iE_{m,7}t/\hbar}$\tabularnewline
$+\cos^{2}\frac{\theta_{1}}{2}\cos^{2}\frac{\theta_{2}}{2}\cos\Theta e^{-iE_{m,10}t/\hbar}+\cos^{2}\frac{\theta_{1}}{2}\sin^{2}\frac{\theta_{2}}{2}\cos\Theta e^{-iE_{m,11}t/\hbar}]$\tabularnewline
\end{tabular}\tabularnewline
\hline
$D_{m,2,3}$ & %
\begin{tabular}{c}
$\frac{1}{8}[-\sin\theta_{1}\sin\theta_{2}\cos\Theta e^{-iE_{m,6}t/\hbar}+\sin\theta_{1}\sin\theta_{2}\cos\Theta e^{-iE_{m,7}t/\hbar}$\tabularnewline
$+\sin\theta_{1}\sin\theta_{2}\cos\Theta e^{-iE_{m,10}t/\hbar}-\sin\theta_{1}\sin\theta_{2}\cos\Theta e^{-iE_{m,11}t/\hbar}]$\tabularnewline
\end{tabular}\tabularnewline
\hline
$D_{m,2,4}$ & $\frac{1}{2}[\cos\frac{\theta_{1}}{2}\sin\frac{\theta_{1}}{2}\sin\Theta e^{-iE_{m,8}t/\hbar}-\sin\frac{\theta_{1}}{2}\cos\frac{\theta_{1}}{2}\sin\Theta e^{-iE_{m,12}t/\hbar}]$\tabularnewline
\hline
$D_{m,3,1}$ & $\frac{1}{2}[\sin\frac{\theta_{1}}{2}\cos\frac{\theta_{1}}{2}\sin\Theta e^{-iE_{m,5}t/\hbar}-\cos\frac{\theta_{1}}{2}\sin\frac{\theta_{1}}{2}\sin\Theta e^{-iE_{m,9}t/\hbar}]$\tabularnewline
\hline
$D_{m,3,2}$ & %
\begin{tabular}{c}
$\frac{1}{8}[\sin\theta_{1}\sin\theta_{2}\cos\Theta e^{-iE_{m,6}t/\hbar}-\sin\theta_{1}\sin\theta_{2}\cos\Theta e^{-iE_{m,7}t/\hbar}$\tabularnewline
$-\sin\theta_{1}\sin\theta_{2}\cos\Theta e^{-iE_{m,10}t/\hbar}+\sin\theta_{1}\sin\theta_{2}\cos\Theta e^{-iE_{m,11}t/\hbar}]$\tabularnewline
\end{tabular}\tabularnewline
\hline
$D_{m,3,3}$ & %
\begin{tabular}{c}
$\frac{1}{2}[-\cos^{2}\frac{\theta_{1}}{2}\sin^{2}\frac{\theta_{2}}{2}\cos\Theta e^{-iE_{m,6}t/\hbar}-\cos^{2}\frac{\theta_{1}}{2}\cos^{2}\frac{\theta_{2}}{2}\cos\Theta e^{-iE_{m,7}t/\hbar}$\tabularnewline
$-\sin^{2}\frac{\theta_{1}}{2}\sin^{2}\frac{\theta_{2}}{2}\cos\Theta e^{-iE_{m,10}(\varphi_{10})t/\hbar}-\sin^{2}\frac{\theta_{1}}{2}\cos^{2}\frac{\theta_{2}}{2}\cos\Theta e^{-iE_{m,11}t/\hbar}+\cos\Theta e^{-iE_{m,13}t/\hbar}]$\tabularnewline
\end{tabular}\tabularnewline
\hline
$D_{m,3,4}$ & $\frac{1}{2}[-\cos\frac{\theta_{2}}{2}\sin\frac{\theta_{2}}{2}\sin\Theta e^{-iE_{m,14}(\varphi_{14})t/\hbar}+\sin\frac{\theta_{2}}{2}\cos\frac{\theta_{2}}{2}\sin\Theta e^{-iE_{m,15}(\varphi_{15})t/\hbar}]$\tabularnewline
\hline
$D_{m,4,1}$ & $0$\tabularnewline
\hline
$D_{m,4,2}$ & %
\begin{tabular}{c}
$\frac{1}{4}[\sin\theta_{1}\cos^{2}\frac{\theta_{2}}{2}\sin\Theta e^{-iE_{m,6}t/\hbar}+\sin\theta_{1}\sin^{2}\frac{\theta_{2}}{2}\sin\Theta e^{-iE_{m,7}t/\hbar}$\tabularnewline
$-\sin\theta_{1}\cos^{2}\frac{\theta_{2}}{2}\sin\Theta e^{-iE_{m,10}t/\hbar}-\sin\theta_{1}\sin^{2}\frac{\theta_{2}}{2}\sin\Theta e^{-iE_{m,11}t/\hbar}]$\tabularnewline
\end{tabular}\tabularnewline
\hline
$D_{m,4,3}$ & %
\begin{tabular}{c}
$\frac{1}{4}[-\cos^{2}\frac{\theta_{1}}{2}\sin\theta_{2}\sin\Theta e^{-iE_{m,6}t/\hbar}+\cos^{2}\frac{\theta_{1}}{2}\sin\theta_{2}\sin\Theta e^{-iE_{m,7}t/\hbar}$\tabularnewline
$-\sin^{2}\frac{\theta_{1}}{2}\sin\theta_{2}\sin\Theta e^{-iE_{m,10}t/\hbar}+\sin^{2}\frac{\theta_{1}}{2}\sin\theta_{2}\sin\Theta e^{-iE_{m,11}t/\hbar}]$\tabularnewline
\end{tabular}\tabularnewline
\hline
$D_{m,4,4}$ & %
\begin{tabular}{c}
$\frac{1}{2}[-\cos\frac{\theta_{1}}{2}\cos\frac{\theta_{1}}{2}\cos\Theta e^{-iE_{m,8}t/\hbar}-\sin\frac{\theta_{1}}{2}\sin\frac{\theta_{1}}{2}\cos\Theta e^{-iE_{m,12}t/\hbar}$\tabularnewline
$+\cos\frac{\theta_{2}}{2}\cos\frac{\theta_{2}}{2}\cos\Theta e^{-iE_{m,14}t/\hbar}+\sin\frac{\theta_{2}}{2}\sin\frac{\theta_{2}}{2}\cos\Theta e^{-iE_{m,15}t/\hbar}]$\tabularnewline
\end{tabular}\tabularnewline
\hline
\end{tabular}

\caption{Coefficients for the probability from the triplet to the singlet state
at time $t$.}
\end{table*}

The total Hamiltonian $\tilde{H}=\tilde{H}^{(0)}+\tilde{H}^{(1)}$
is split into two parts, i.e., $\tilde{H}^{(0)}=\tilde{H}_{r}^{(0)}+\tilde{H}_{s}$
governing the free dynamic evolution and $\tilde{H}^{(1)}=\tilde{H}_{r}^{(1)}$
for the ET process. $|\psi_{mq}^{(0)}\rangle=|\varphi_{q}\rangle\otimes|m\rangle$
are the eigenstates of $\tilde{H}^{(0)}$ with eigenvalues $E_{mq}^{(0)}=m\hbar\omega+E_{sq}$.
Here, $m\ (m=0,...,\infty)$ denotes the relative vibrational quantum
number.A given initial state is expanded as
\begin{equation}
|\Psi_{m}(0)\rangle=\sum_{q=1}^{24}c_{mq}(0)|\psi_{mq}^{(0)}\rangle,
\end{equation}
 and then the wave function at time $t$ is given by
\begin{equation}
|\Psi(t)\rangle=\sum_{n=0}^{\infty}\sum_{p=1}^{24}c_{np}(t)\exp[-iE_{np}^{(0)}t/\hbar]|\psi_{np}^{(0)}\rangle,
\end{equation}
where $c_{np}(t)$ are the coefficients determined by the Schr$\ddot{\mbox{o}}$dinger
equation

\begin{equation}
i\hbar\dot{c}_{np}(t)=\sum_{n',p'}e^{i\omega_{np,n'p'}t}\tilde{H}_{np,n'p'}^{(1)}c_{n',p'}(t).\label{eq:A1}
\end{equation}
 Here,
\begin{equation}
\omega_{np,n'p'}=[E_{np}^{(0)}-E_{n'p'}^{(0)}]/\hbar,
\end{equation}
 and
\begin{equation}
\tilde{H}_{np,n'p'}^{(1)}=\langle\psi_{np}^{(0)}|H^{(1)}|\psi_{n'p'}^{(0)}\rangle.
\end{equation}

To the first-order approximation, as $c_{n'p'}(t)$ in the right-hand
side of Eq.~(\ref{eq:A1}) is approximated as $c_{mp'}(0)\delta_{n',m}$,
it is straightforward to obtain

\begin{equation}
c_{np}(\tau)=c_{np}(0)-\sum_{q}c_{mq}(0)\tilde{H}_{np,mq}^{(1)}\frac{e^{i\omega_{np,mq}\tau}-1}{\hbar\omega_{np,mq}}.
\end{equation}
The square of its norm gives the probability $P_{np}$ of finding
the system in the state $|\psi_{np}^{(0)}\rangle$ at time $\tau$.
Thus, for $(n,p)\neq(m,q)$, we have
\begin{equation}
P_{np}(\tau)=\left|\sum_{q}c_{mq}(0)\tilde{H}_{np,mq}^{(1)}\frac{e^{i\omega_{np,mq}\tau}-1}{\hbar\omega_{np,mq}}\right|^{2}.\label{eq:A2}
\end{equation}
The total ET reaction probability is $P(\tau)=\sum_{np}P_{np}(\tau)$.
And the chemical reaction rate is determined by the reaction probability
per unit time in the long-time limit~\cite{Sakurai}, i.e.,
\begin{equation}
k=\frac{\partial}{\partial\tau}\lim_{\tau\rightarrow\infty}P(\tau).
\end{equation}

In our case, the system is initially in the state $\rho(0)=\rho_{v}\otimes\rho_{n}\otimes\rho_{t}$,
i.e., the relative vibration part
\begin{equation}
\rho_{v}=\frac{1}{Z}\sum_{m=0}^{\infty}e^{-\beta m\hbar\omega}\left|m\right\rangle \left\langle m\right|,
\end{equation}
 with
\begin{equation}
Z=1/[1-\exp(-\hbar\omega/k_{B}T)],
\end{equation}
the nuclear spin part
\begin{equation}
\rho_{n}=\frac{1}{4}\sum_{j=1}^{4}\left|\chi_{j}^{n}\right\rangle \left\langle \chi_{j}^{n}\right|,
\end{equation}
 and the electron spin part
\begin{equation}
\rho_{t}=\left|t\right\rangle \left\langle t\right|.
\end{equation}

After the Van Vleck transformation (\ref{van Vleck transformation})
and the rotation for the system, the initial state is transformed
as $\tilde{\rho}(0)=\tilde{\rho}_{v}\otimes\tilde{\rho}_{n}\otimes\tilde{\rho}_{t}$,
where $\tilde{\rho}_{v}=\rho_{v}$, $\tilde{\rho}_{n}=\rho_{n}$,
and $\tilde{\rho}_{t}=|\tilde{t}\rangle\langle\tilde{t}|$. And we
calculate the total chemical reaction probability as
\begin{equation}
P_{t}(\tau)=\frac{1}{4Z}\sum_{m,n=0}^{\infty}\sum_{j=1}^{4}\sum_{p=1}^{24}e^{-\beta m\hbar\omega}P_{jmnp}(\tau),
\end{equation}
 where
\begin{eqnarray}
P_{jmnp}(\tau) & = & \left|\sum_{q}c_{jmq}\tilde{H}_{np,mq}^{(1)}\frac{1-e^{i\omega_{np,mq}\tau}}{\hbar\omega_{np,mq}}\right|^{2},\label{reaction probability}\\
c_{jmq} & = & \langle\psi_{mq}^{(0)}|\tilde{t}\rangle|\chi_{j}^{n}\rangle|m\rangle.
\end{eqnarray}
As a matter of fact, those energy-increasing terms with $q=17,18,19,20$
(corresponding to the final states with both electrons at the orbits
of the donor) and the cross-product terms with $\omega_{np,mq}\neq0$
of Eq.~(\ref{reaction probability}) do not contribute much to the
chemical reaction probability. Therefore, a Fermi's golden-rule-like
chemical reaction rate is obtained as
\begin{equation}
k_{t}=\frac{\pi J^{2}}{4Z\hbar}\sum_{m,n=0}^{\infty}\sum_{j=1}^{4}\sum_{p=21}^{24}e^{-\beta m\hbar\omega}|\langle n|e^{-2\phi(b^{\dagger}-b)}|m\rangle|^{2}R_{jmnp}
\end{equation}
with the coefficients being listed in Table~II.

\section{Singlet and triplet states interconversion}

For the system initially in the state
\begin{equation}
\tilde{\rho}(0)=\tilde{\rho}_{v}\otimes\tilde{\rho}_{n}\otimes\tilde{\rho}_{t},
\end{equation}
the probability of the electrons converted to the singlet state at
time $t$ reads
\begin{eqnarray}
P_{t\rightarrow s} & = & \mathrm{Tr}_{v,n,e}\left[\tilde{\rho}_{s}e^{-i\tilde{H}t/\hbar}\tilde{\rho}(0)e^{i\tilde{H}t/\hbar}\right]\\
 & = & \frac{1}{4Z}\sum_{m,n}\sum_{i,j}e^{-\beta m\hbar\omega}\nonumber \\
 &  & \times\left|\langle\tilde{s}|\langle\chi_{i}^{n}|\langle n|e^{-i\tilde{H}t/\hbar}|m\rangle|\chi_{j}^{n}\rangle|\tilde{t}\rangle\right|^{2}
\end{eqnarray}
The eigenfunction of $\tilde{H}$ is approximated to the first order
as
\begin{eqnarray}
\left|\psi_{mq}\right\rangle  & = & |\psi_{mq}^{(0)}\rangle+\sum_{n=0}^{\infty}\sum_{p=1}^{24}\xi(m,n,q,p)|\psi_{np}^{(0)}\rangle
\end{eqnarray}
 with
\begin{eqnarray}
\xi(m,n,q,p) & = & \frac{\langle\psi_{np}^{(0)}|\tilde{H}^{(1)}|\psi_{mq}^{(0)}\rangle}{E_{mq}^{\left(0\right)}-E_{np}^{\left(0\right)}},\label{eq:B2}
\end{eqnarray}
while the eigen energy is obtained to the second order as
\begin{eqnarray}
E_{mq} & = & E_{mq}^{(0)}+\sum_{n,p}\frac{|\langle\psi_{np}^{(0)}|\tilde{H}^{(1)}|\psi_{mq}^{(0)}\rangle|^{2}}{E_{mq}^{\left(0\right)}-E_{np}^{\left(0\right)}}.
\end{eqnarray}
 As a result, the time evolution operator is approximated as
\begin{eqnarray}
e^{-i\tilde{H}t/\hbar} & = & \sum_{k=0}^{\infty}\sum_{q=1}^{24}e^{\pm iE_{kq}t/\hbar}\left|\psi_{kq}\right\rangle \left\langle \psi_{kq}\right|.
\end{eqnarray}
 Neglecting the the second-order terms, we obtain the conversion probability
as
\begin{eqnarray}
P_{t\rightarrow s} & = & \frac{1}{4Z}\sum_{m=0}^{\infty}\sum_{j,k=1}^{4}e^{-\beta m\hbar\omega}\nonumber \\
 &  & \times\left|\sum_{q=1}^{16}\langle\check{t}|\langle\chi_{k}^{n}|\varphi_{q}\rangle\langle\varphi_{q}|\chi_{j}^{n}\rangle|\check{s}\rangle e^{-iE_{mq}t/\hbar}\right|^{2}\nonumber \\
 & = & \frac{1}{4Z}\sum_{m=0}^{\infty}\sum_{j,k=1}^{4}e^{-\beta m\hbar\omega}|D_{mjk}|^{2},
\end{eqnarray}
 where the coefficients $D_{mjk}$ are listed in Table~III.

\end{document}